\documentclass[11pt]{article}
\usepackage[dvips]{color}
\usepackage{epsfig}
\usepackage{amsmath}
\usepackage{graphicx}
\date{}

\begin{document}
\title{\bf Classical and quantum Chaplygin gas Ho\v{r}ava-Lifshitz scalar-metric cosmology}
\author{H. Ardehali$^1$,\,\, P. Pedram$^1$\thanks{%
p.pedram@srbiau.ir},\,\, and B. Vakili$^2$\thanks{%
b.vakili@iauctb.ac.ir}
\\\\
$^1${\small {\it Department of Physics, Science and Research Branch, Islamic Azad University, Tehran, Iran}} \\
$^2${\small {\it Department of Physics, Central Tehran Branch,
Islamic Azad University, Tehran, Iran}}}

\maketitle

\begin{abstract}
In this work, we study the Friedmann-Robertson-Walker cosmology in
which a Chaplygin gas is coupled to a non-linear scalar field in the
framework of the Ho\v{r}ava-Lifshitz theory. In writing the action
of the matter part, we use the Schutz's formalism so that the only
degree of freedom of the Chaplygin gas plays the role of an
evolutionary parameter. In a minisuperspace perspective, we
construct the Lagrangian for this model and show that in comparison
with the usual Einstein-Hilbert gravity, there are some correction
terms coming from the Ho\v{r}ava theory. In such a set-up and by
using of some approximations the classical dynamics of the model is
investigated and some discussions about their possible singularities
are presented. We then deal with the quantization of the model in
the context of the Wheeler-DeWitt approach of quantum cosmology to
find the cosmological wave function. We use the resulting wave
functions to investigate the possibility of the avoidance of
classical singularities due to quantum effects.
\vspace{5mm}\noindent\vspace{0.8mm}\newline

\end{abstract}

\section{Introduction}\label{sec_Introduction}
Various modern cosmological theories such as grand unified theories
imply the existence of the classical and semiclassical scalar fields
\cite{SF1}. In cosmological viewpoint, scalar-tensor models have
been attracted much attention in which a non-minimal coupling
appears between the space-time geometry and a scalar field
\cite{SF2,SF3,SF4,SF5}. This is due to the fact that various
research areas in cosmology such as spatially flat and accelerated
expanding universe at the present time \cite{SF6,SF7,SF8}, inflation
\cite{inflation1,inflation2}, dark matter and dark energy
\cite{dark1,dark2}, and many other behaviors can be explained
phenomenologically by the scalar fields. Cosmological models usually
described by a single scalar field with a canonical kinetic term in
the form $\frac{1}{2}g^{\mu\nu}\partial_{\mu}\phi\partial_{\nu}\phi$
and a self-interaction potential $V(\phi)$ where the scalar field is
often minimally coupled to gravity. However, in scalar-tensor
theories, the scalar field is not simply added to the action.
Indeed, it is added to the tensor gravitational field by a
non-minimal coupling term \cite{non-minimalcoupling}.

In recent years, the so-called Ho\v{r}ava-Lifshitz (HL) gravity
theory, presented by Ho\v{r}ava, is proved to be power-countable
renormalizable. It is based on the anisotropic scaling of space
$\mathbf{x}$ and time $t$ as
\begin{eqnarray}\label{asymmetry_scaling}
\mathbf{x}\rightarrow b\mathbf{x},\qquad t\rightarrow b^zt,
\end{eqnarray}
where $b$ is a scaling parameter and $z$ is the dynamical critical
exponent. Notice that for $z = 1$ the standard relativistic scale
invariance obeying Lorentz symmetry is recovered in the IR limit.
However, the UV gravitational theory implies $z = 3$
\cite{Horava1,Horava2,Horava3,Horava4}. Due to the asymmetry of
space and time in HL theory, it is common to use the
Arnowitt-Deser-Misner (ADM) formalism to represent the space-time
metric $g_{\mu\nu}(t,\mathbf{x})$, in terms of three-dimensional
metric $\gamma_{ab}(t,\mathbf{x})$, shift vector $N_a(t,\mathbf{x})$
and the lapse function $N(t,{\bf x})$ as \cite{ADM}
\begin{equation}
g_{\mu \nu}(t,{\bf x})=\left(%
\begin{array}{cc}
-N^2(t,{\bf x})+N_a(t,{\bf x})N^a(t,{\bf x}) & N_b(t,{\bf x}) \\
N_a(t,{\bf x}) & \gamma_{ab}(t,{\bf x}) \\
\end{array}%
\right).
\end{equation}
If the lapse function is a function of $t$ only, the theory is
projectable, otherwise, in the case where $N$ is a function of $(t,
\mathbf{x})$, theory is called non-projectable. General cases in
which the lapse function is taken as a non-projectable function are
studied in Ref.\cite{non-projectable1,non-projectable2}. However,
we assume the lapse function is constrained to be a function only of
the time coordinate $N=N(t)$ \cite{Horava2}.

The most general action for HL gravity (without the detailed balance
condition) is given by $S_{HL}=S_K+S_V$, where $S_K$ is kinetic part
\begin{eqnarray}\label{S_K}
S_K \sim \int d^4\mathbf{x}\;\sqrt{-g}(K_{ij}K^{ij}-\lambda K^2),
\end{eqnarray}
in which $K_{ij}$ is the extrinsic curvature tensor (with trace $K$)
defined by
\begin{eqnarray}\label{K_ij}
K_{ij}=\frac{1}{2N}\left(\dot{\gamma}_{ij}-\nabla_iN_j-\nabla_jN_i\right).
\end{eqnarray}
Also, for the potential part the following general form is proposed
\begin{eqnarray}\label{S_V}
S_{V}=\int d^4x\;\sqrt{-g}V[\gamma_{ij}],
\end{eqnarray}
in which
\begin{eqnarray}\label{V}
V[\gamma_{ij}]&=&g_0\zeta^6+g_1\zeta^4R+g_2\zeta^2R^2+g_3\zeta^2R_{ij}R^{ij}\nonumber\\
&&+g_4R^3+g_5RR_{ij}R^{ij}+g_6R^i_{\,\,j}R^j_{\,\,k}R^k_{\,\,i}\nonumber\\
&&+g_7R\nabla^2R+g_8\nabla_iR_{jk}\nabla^iR^{jk}.
\end{eqnarray}The constants $\lambda$ and $g_i$
($i=0,1,...,8$) in above relations, denote the HL corrections to the
usual Einstein gravity and $\zeta$ is introduced to make the
constants $g_k$s dimensionless. Under these conditions, the full HL
action that we shall study is
\cite{vakili&kord,Sotiriou,Pitelli&Saa}
\begin{eqnarray}\label{action_HL1}
S_{HL}&=&\frac{M_{PL}^2}{2}\int_\mathcal{M}d^4\mathbf{x}\sqrt{-g}\Big[K_{ij}K^{ij}-\lambda K^2+R-2\Lambda\nonumber\\
&&\qquad-\frac{g_2}{M_{PL}^2}R^2-\frac{g_3}{M_{PL}^2}R_{ij}R^{ij}-\frac{g_4}{M_{PL}^4}R^3
-\frac{g_5}{M_{PL}^4}RR_{ij}R^{ij}\nonumber\\
&&\qquad-\frac{g_6}{M_{PL}^4}R_{ij}R^{jk}R^i_{\;k}
-\frac{g_7}{M_{PL}^4}R\nabla^2R-\frac{g_8}{M_{PL}^4}\nabla_iR_{jk}\nabla^iR^{jk}\Big],
\end{eqnarray}in which $M_{PL}=\frac{1}{\sqrt{8\pi G}}$ and we have
set $c=1$, $\zeta=1$, $\Lambda=g_0M_{Pl}^2/2$ and $g_1=-1$.

All cosmological evidences have revealed that the universe is
undergoing an accelerated expansion which can be described by exotic
cosmic fluid, the so-called dark energy, one of the first model of
which is the cosmological constant. On the other hand, scalar fields
play an important role in unified theories of interactions and also
in inflationary scenarios in cosmology. Indeed, a rich variety of
dark energy and inflationary models can be accommodated
phenomenologically by scalar fields in which the inflatons produce
the initial acceleration. Another attempt, originally raised in
string theory \cite{John1}, is to change the equation of state from
an ordinary matter to the Chaplygin gas, an exotic fluid with
negative pressure. Chaplygin gas as a candidate behind the current
observation of cosmic acceleration has been thoroughly investigated
in recent years. The generalized Chaplygin gas with negative
pressure is described by an exotic equation of state
\begin{eqnarray}\label{CG_equation_of_atate}
P=-\frac{A}{\rho^\alpha},
\end{eqnarray}
where $P$ is the pressure, $A$ is a positive constant, and
$0\leq\alpha\leq1$ is the equation of state parameter such that
$\alpha=1$ denotes the standard Chaplygin gas \cite{CG1,CG2}. In
this sense, since string theory deals with the high energy phenomena
such as very early universe, considering the chaplygin gas quantum
cosmology may have physical grounds. It is shown that
\cite{Herrera}, the generalized Chaplygin gas
(\ref{CG_equation_of_atate}) can play the role of a mixture of
cosmological constant and radiation by means of which the the
cosmological dynamics shows a transition from a dust dominated era
to a de Sitter phase and thus it interpolates between dust matter
and the cosmological constant. Cosmology with generalized Chaplygin
gas (\ref{CG_equation_of_atate}) results in an expanding universe
which begins from a non-relativistic matter dominated phase and ends
at a cosmological constant dominated era \cite{CG2}. Also, the idea
of this fluid is used to find a solution to the coincidence problem
in cosmology \cite{CG3,CG4,CG5,CG6,CG8,CG9,CG10}. Quantum
cosmological models with Chaplygin gas have been studied in
Refs.~\cite{CG7,Pedram&Jalalzadeh&Gousheh,ardehali&pedram},
specially in Ref.~\cite{Majumder}, a scalar field is also added to
the Chaplygin gas quantum cosmology and its effects are
investigated. In summary, since the Chaplygin gas models are able to
describe the smooth transition from a decelerated expansion to an
accelerated universe and also since they try to give a unified
picture of dark matter and dark energy, one may use them as an
alternative to the traditional $\Lambda$CDM models.

In this paper we shall consider a cosmological model in the
framework of a projectable HL gravity without detailed balance
condition. A Chaplygin gas will play the role of the matter source
and a scalar field is coupled to metric with a generic coupling
function $F(\phi)$. The classical version of such models are used to
answer the missing-matter problem in cosmology \cite{Saez&Ballester}
and their quantum cosmology is studied in
Refs.~\cite{Majumder,vakili,Socorro&Sabido&Urena-Lopez}. Since our
aim in the quantum part of the model is to investigate the time
evolution of the wave function, we prefer to use the Chaplygin gas
in the framework of the Schutz formalism \cite{schutz1,schutz2}. In
such a setup the Hamiltonian of the gas consists of a linear
momentum, the variable canonically conjugate to which may play the
role of a time parameter (see
Refs.~\cite{vakili&kord,Pedram&Jalalzadeh&Gousheh,ardehali&pedram,vakili,Lapchinskii&Rubakov,Bertolami&Zarro}
for details of this formalism).

The paper in organized as follows: In
Sec.~\ref{Sec_ConstructionHamiltonian}, we construct the action of
HL gravity with Chaplygin gas and scalar field in terms of
minisuperspace variables. In Secs.~\ref{Sec_Limit1} and
\ref{Sec_Limit2}, we approximate the super-Hamiltonian in two cases
$Sp_{\epsilon}^{\alpha+1}\gg Aa^{3(\alpha+1)}$ and
$Sp_{\epsilon}^{\alpha+1}\ll Aa^{3(\alpha+1)}$ separately. Schutz
formalism for Chaplygin gas allows us to get a
Schr\"{o}dinger-Wheeler-DeWitt (SWD) equation in which the only
remaining matter degree of freedom plays the role of time. After
choosing the coupling function between the scalar field and metric
as $F(\phi)=\lambda\phi^m$, we obtain the classical dynamics of the
scale factor and scalar field in terms of the Schutz's time
parameter and see that they exhibit some types of singularities. We
then deal with the quantization of the model and by computing the
expectation values of the scale factor and scalar field we show that
the evolution of the universe based on the quantum picture is free
of classical singularities. Section \ref{sec_conclusion} is devoted
to summary and conclusions.

\section{The model}\label{Sec_ConstructionHamiltonian}
The total action (without the detailed balance condition) of our
model consists of three parts, that are, gravitational
Ho\v{r}ava-Lifshitz gravity action, scalar field and Chaplygin gas
actions parts as
\begin{eqnarray}\label{total_action}
S=S_{HL}+S_{\phi}+S_P,
\end{eqnarray}
where $S_{HL}$, $S_{\phi}$ and $S_P$ are the Ho\v{r}ava-Lifshitz,
scalar field and Chaplygin gas actions, respectively. Now, we expand
them separately.

\subsection{Ho\v{r}ava-Lifshitz action}\label{subsec_HL_gravity}
The action for the projectable HL gravity without detailed balance
is given in (\ref{action_HL1}). In a quasi-spherical polar
coordinate system, we assume that the geometry of space-time is
described by the FRW metric
\begin{eqnarray}\label{metricFRW}
ds^2&=&g_{\mu\nu}dx^{\mu}dx^{\nu}\nonumber\\
&=&-N^2(t)dt^2+a^2(t)\left[\frac{dr^2}{1-kr^2}+r^2\left(d\vartheta^2+\sin^2\vartheta
d\varphi\right)\right],
\end{eqnarray}
in which $N(t)$ is the lapse function, $a(t)$ is the scale factor
and $k=-1,0,+1$ denotes the open, flat, and closed universes,
respectively. Now, in the language of the ADM variables the above
metric can be rewritten as
\[ds^2=-N^2(t)dt^2+\gamma_{ij}dx^idx^j,\] where
\begin{eqnarray}\label{3-metric}
\gamma_{ij}=a^2(t)\mathrm{diag}\left(\frac{1}{1-kr^2},r^2,r^2\sin^2\vartheta\right),
\end{eqnarray}is
the induced intrinsic metric on the $3$-dimensional spatial
hypersurfaces from which we obtain the Ricci and extrinsic curvature
tensors as
\begin{eqnarray}\label{R&K}
R_{ij}=\frac{2k}{a^2}\gamma_{ij},\qquad
K_{ij}=\frac{\dot{a}}{Na}\gamma_{ij}.
\end{eqnarray}The
gravitational part for the model may now be written by substituting
the above results into action (\ref{action_HL1}) giving
\begin{eqnarray}\label{action_HL2}
S_{HL}&=&\frac{3(3\lambda-1)M^2_{PL}V_0}{2}\int
dt\;Na^3\left[-\frac{\dot{a}^2}{N^2a^2}
+\frac{6k}{3(3\lambda-1)}\frac{1}{a^2}\right.\nonumber\\
&&\quad\left.-\frac{2\Lambda}{3(3\lambda-1)}-\frac{12k^2}{a^4}\frac{3g_2+g_3}{3(3\lambda-1)M^2_{PL}}
-\frac{24k^3}{a^6}\frac{9g_4+3g_5+g_6}{3(3\lambda-1)M^4_{PL}}\right]\nonumber\\
&=&\int
dt\;N\left(-\frac{a\dot{a}^2}{N^2}+g_ca-g_{\Lambda}a^3-\frac{g_r}{a}-\frac{g_s}{a^3}\right),
\end{eqnarray}
where $V_0=\int d^3x\frac{r^2\sin\vartheta}{\sqrt{1-kr^2}}$ is the
integral over spatial dimensions. Also, we have defined the
coefficients $g_c$, $g_{\Lambda}$, $g_r$ and $g_s$ as

\begin{eqnarray}\label{gi}
\left\{
\begin{array}{ll}
&g_c=\frac{6k}{3(3\lambda-1)},\\
&g_{\Lambda}=\frac{2\Lambda}{3(3\lambda-1)},\\
&g_r=\frac{12k^2(3g_2+g_3)}{3(3\lambda-1)M^2_{PL}},\\
&g_s=\frac{24k^3(9g_4+3g_5+g_6)}{3(3\lambda-1)M^4_{PL}},
\end{array}\right.
\end{eqnarray}in which we have set $3V_0
M_{Pl}^2(3\lambda-1)/2=1$. Now, the gravitational part of the
Hamiltonian for this model can be obtained from its standard
procedure. Noting that
\begin{eqnarray}\label{momenta_HL}
p_a=-\frac{2a\dot{a}}{N},
\end{eqnarray}
we get
\begin{eqnarray}\label{hamiltonian_HL}
H_{HL}&=&p_a\,\dot{a}-\mathcal{L}_{HL},\nonumber\\
&=&N\left(-\frac{p_a^2}{4a}-g_ca+g_{\Lambda}a^3+\frac{g_r}{a}+\frac{g_s}{a^3}\right).
\end{eqnarray}

\subsection{The Chaplygin gas}\label{subsec_CG}
In Schutz formalism, the four velocity of a fluid can be expressed
in terms of six scalar potentials as \cite{schutz1,schutz2}
\begin{eqnarray}\label{Schutz_expresstion}
u_{\nu}=\frac{1}{\mu}(\partial_\nu\epsilon+\varpi\partial_\nu\beta+\theta\partial_\nu
S),
\end{eqnarray}
where $\mu$ and $S$ are specific enthalpy and entropy respectively
while the potentials $\varpi$ and $\beta$ are related to torsion and
are absent in FRW models. The potentials $\epsilon$ and $\theta$
have no direct physical interpretation in this formalism. The
four-velocity obeys the condition $u_{\nu}u^{\nu}=1$. Hence, the
four-velocity of the fluid in its rest frame reads
\begin{eqnarray}\label{Schutz_expresstion2}
u_{\nu}=N\delta^0_{\nu}\quad\Rightarrow\mu=\frac{\dot{\epsilon}+\theta\dot{S}}{N}.
\end{eqnarray}
Following the thermodynamical description of
\cite{Pedram&Jalalzadeh&Gousheh,Lapchinskii&Rubakov}, the basic
thermodynamic relations of the Chaplygin gas are given by
\begin{eqnarray}\label{thermodynamic_CG}
\rho=\rho_0(1+\Pi),\qquad\mu=1+\Pi+\frac{P}{\rho_0},
\end{eqnarray}
where $\rho_0$ and $\Pi$ are the rest mass density and  the specific
internal energy of the gas, respectively. These quantities together
with the temperature $\tau$ of the system obey the first law of the
thermodynamics, which can be rewritten as
\begin{eqnarray}\label{first_law_thermodynamics}
\tau dS&=&d\Pi+Pd\left(\frac{1}{\rho_0}\right)\nonumber \\
&=&\frac{1}{(1+\alpha)(1+\Pi)^{\alpha}}d\left[(1+\Pi)^{1+\alpha}-\frac{A}{\rho_0^{1+\alpha}}\right],
\end{eqnarray}where we have used the equation of state
(\ref{CG_equation_of_atate}). Therefore, the temperature and entropy
of the gas are obtained as
\begin{eqnarray}\label{tau&S}
\tau=\frac{1}{(1+\alpha)(1+\Pi)^{\alpha}},\qquad
S=(1+\Pi)^{1+\alpha}-\frac{A}{\rho_0^{1+\alpha}}.
\end{eqnarray}
Now, we can express the energy density and pressure as
functions of $\mu$ and $S$
\begin{eqnarray}\label{rho}
\rho&=&\left[\frac{1}{A}\left(1-\frac{\mu^{\frac{\alpha+1}{\alpha}}}{S^{\frac{1}{\alpha}}}\right)\right]
^{\frac{-1}{\alpha+1}},\\\label{P}
P&=&-A\left[\frac{1}{A}\left(1-\frac{\mu^{\frac{\alpha+1}{\alpha}}}{S^{\frac{1}{\alpha}}}\right)\right]
^{\frac{\alpha}{\alpha+1}}.
\end{eqnarray}
Finally, with the help of these relations, the action of the
Chaplygin gas takes the form
\begin{eqnarray}\label{action_CG}
S_{P}&=&\int dtd^3\mathrm{x}\,N\sqrt{\gamma}P\nonumber \\  &=&-A\int
dt\,Na^3\left[\frac{1}{A}\left(1-\frac{(\dot{\epsilon}+\theta\dot{S})^{\frac{\alpha+1}{\alpha}}}
{N^{\frac{\alpha+1}{\alpha}}S^{\frac{1}{\alpha}}}\right)\right]^{\frac{\alpha}{\alpha+1}}.
\end{eqnarray}
Now, in terms of the conjugate momenta
\begin{eqnarray}\label{momenta_CG}
\begin{array}{ll}
&p_a=p_{\theta}=0,\\
&p_{\epsilon}=a^3\left(\frac{\dot{\epsilon}+\theta\dot{S}}{NS}\right)^{\frac{1}{\alpha}}
\left[\frac{1}{A}\left(1-\frac{(\dot{\epsilon}+\theta\dot{S})^{\frac{\alpha+1}{\alpha}}}
{N^{\frac{\alpha+1}{\alpha}}S^{\frac{1}{\alpha}}}\right)\right]^{\frac{-1}{\alpha+1}},\\
&p_{S}=\theta p_{\epsilon},
\end{array}
\end{eqnarray}
the Chaplygin gas Hamiltonian can be written as follows
\begin{eqnarray}\label{hamiltonian_CG1}
H_P&=&(\dot{\epsilon}+\theta\dot{S})p_{\epsilon}-\mathcal{L}_{P}\nonumber\\
&=&Na^3\left[\frac{1}{A}\left(1-\frac{(\dot{\epsilon}+\theta\dot{S})^{\frac{\alpha+1}{\alpha}}}
{N^{\frac{\alpha+1}{\alpha}}S^{\frac{1}{\alpha}}}\right)\right]^{\frac{-1}{\alpha+1}}\nonumber\\
&=&N\left(Sp_{\epsilon}^{\alpha+1}+Aa^{3(\alpha+1)}\right)^{\frac{1}{\alpha+1}}.
\end{eqnarray}

\subsection{The Scalar field}\label{subsec_SF}
As mentioned before, we consider a non-linear self-coupling scalar
field minimally coupled to geometry by the coupling function
$F(\phi)$ \cite{vakili}. The action of such a scalar field is
\begin{eqnarray}\label{action_SF1}
S_{\phi}=-\frac{M_{PL}^2}{2}\int
d^4x\,\sqrt{-g}\,F(\phi)g^{\mu\nu}\partial_{\mu}\phi\partial_{\nu}\phi,
\end{eqnarray}
where by substituting the metric (\ref{metricFRW}) in which one gets
\begin{eqnarray}\label{action_SF2}
S_{\phi}=\int dt\,\frac{1}{N}F(\phi)a^3\dot{\phi}^2.
\end{eqnarray}
Noting that the momentum congugate to $\phi$ is
\begin{eqnarray}\label{momenta_SF}
p_{\phi}=\frac{2}{N}F(\phi)a^3\dot{\phi},
\end{eqnarray}
the Hamiltonian of the scale field is obtained as
\begin{eqnarray}\label{hamiltonian_SF}
H_{\phi}=\frac{N\,p_{\phi}^2}{4F(\phi)a^3}.
\end{eqnarray}
Now, we are ready to write the total Hamiltonian for our model as
\begin{eqnarray}
H&=&H_{HL}+H_P+H_{\phi}\nonumber\\
&=&N\Big[-\frac{p_a^2}{4a}-g_ca+g_{\Lambda}a^3+\frac{g_r}{a}+\frac{g_s}{a^3}
+\frac{p_{\phi}^2}{4F(\phi)a^3}\nonumber\\\label{full_hamiltonian}
&&\qquad+\left(Sp_{\epsilon}^{\alpha+1}+Aa^{3(\alpha+1)}\right)^{\frac{1}{\alpha+1}}\Big].
\end{eqnarray}
The setup for constructing the phase space and writing the
Lagrangian and Hamiltonian of the model is now complete. However,
the resulting classical (and quantum) equations of motion do not
seem to have analytical solutions. To extract exact solutions, we
first apply some approximation on the above Hamiltonian
\cite{Pedram&Jalalzadeh&Gousheh}, and then will deal with the
behavior of its classical and quantum pictures.

\section{The $Sp_{\epsilon}^{\alpha+1}\gg Aa^{3(\alpha+1)}$ limit}\label{Sec_Limit1}
In the early times of cosmic evolution when the scale factor is
small, we can use the following expansion
\cite{Bouhmadi-Lopez&Moniz,Bouhmadi-Lopez&Gonzalez-Diaz&Martin-Moruno}
\begin{eqnarray}
\left(Sp_{\epsilon}^{\alpha+1}+Aa^{3(\alpha+1)}\right)^{\frac{1}{\alpha+1}}&=&
S^{\frac{1}{\alpha+1}}p_{\epsilon}\left(1+\frac{Aa^{3(\alpha+1)}}{Sp_{\epsilon}^{\alpha+1}}\right)
^{\frac{1}{\alpha+1}}\nonumber\\
&=&S^{\frac{1}{\alpha+1}}p_{\epsilon}\left(1+\frac{1}{\alpha+1}\frac{Aa^{3(\alpha+1)}}{Sp_{\epsilon}
^{\alpha+1}}+\ldots\right)\nonumber\\\label{approximation1}
&\simeq&S^{\frac{1}{\alpha+1}}p_{\epsilon}.
\end{eqnarray}
Therefore, the super-Hamiltonian takes the form
\begin{eqnarray}\label{approximated_hamiltonian1}
H=N\,\left(-\frac{p_a^2}{4a}-g_ca+g_{\Lambda}a^3+\frac{g_r}{a}+\frac{g_s}{a^3}
+\frac{p_{\phi}^2}{4F(\phi)a^3}+S^{\frac{1}{\alpha+1}}p_{\epsilon}\right).
\end{eqnarray}
Now, consider the following canonical transformation
\cite{Pedram&Jalalzadeh&Gousheh,Pedram&Jalalzadeh}
\begin{eqnarray}
\begin{array}{ll}
&T=-(\alpha+1)S^{\frac{\alpha}{\alpha+1}}p_{\epsilon}^{-1}p_S,\\\label{canonical_transformation1}
&p_T=S^{\frac{1}{\alpha+1}}p_{\epsilon},
\end{array}
\end{eqnarray}
under the act of which Hamiltonian (\ref{approximated_hamiltonian1})
takes the form
\begin{eqnarray}\label{hamiltonian1}
H=N\,\left(-\frac{p_a^2}{4a}-g_ca+g_{\Lambda}a^3+\frac{g_r}{a}+\frac{g_s}{a^3}
+\frac{p_{\phi}^2}{4F(\phi)a^3}+p_T\right).
\end{eqnarray}
We see that the momentum $p_T$ is the only remaining canonical
variable associated with the Chaplygin gas and appears linearly in
the Hamiltonian.
\subsection{The classical model}
\begin{figure}[t]
\begin{center}
\includegraphics[width=0.45\textwidth]{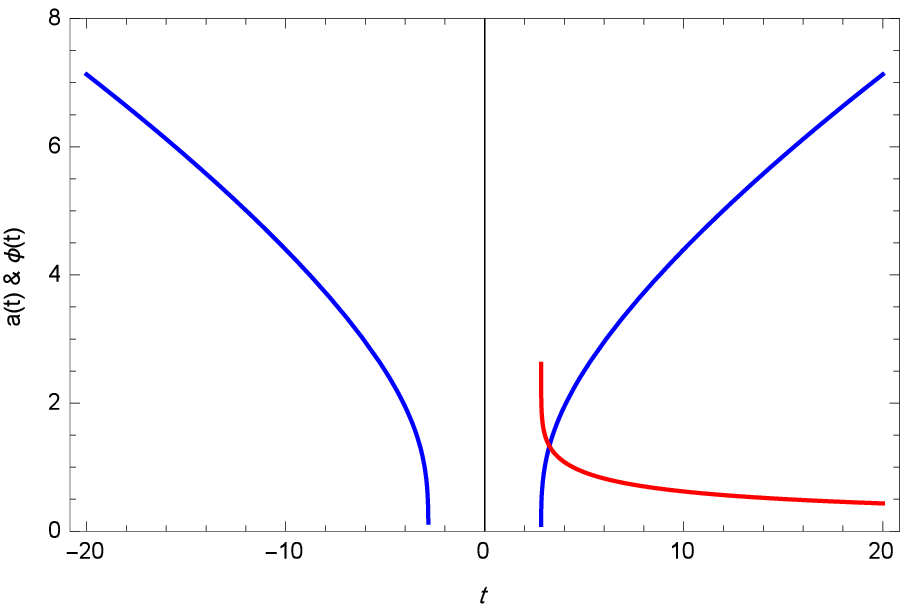}
\includegraphics[width=0.45\textwidth]{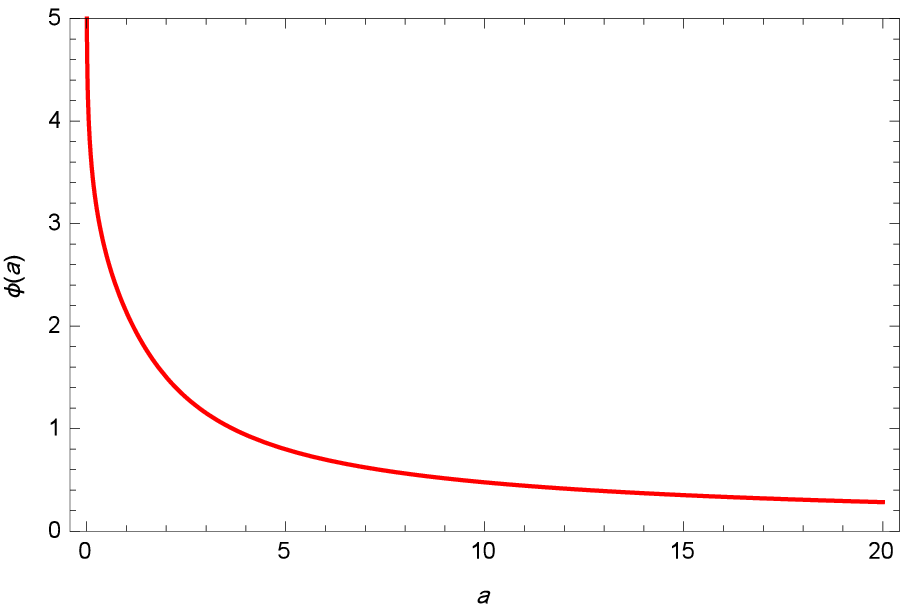}
\caption{Left: The classical scale factor $a(t)$ (blue line) and
$\phi(t)$ (red line). Right: The classical trajectory in $a-\phi$
plane. The figures are plotted for the numerical values
$g_s=\frac{1}{20}$, $p_T=\frac{4}{9\sigma^2}$, $C=3$, $\lambda=1$
and $m=2$.} \label{fig:figure_31}
\end{center}
\end{figure}

The classical dynamics of the system is governed by the Hamiltonian
equation of motion $\dot{q}=\{q,H\}$, for each variable. The result
is
\begin{eqnarray}\label{classical dynamics1}
\left\{
\begin{array}{ll}
\dot{a}=\frac{Np_a}{2a},\\
\dot{p}_a=N\left(-\frac{p_a^2}{4a^2}+g_c-3g_{\Lambda}a^2+\frac{g_r}{a^2}+\frac{3g_s}{a^3}
+\frac{3p_{\phi}^2}{4Fa^4}\right),\\
\dot{\phi}=\frac{Np_{\phi}}{2Fa^3},\\
\dot{p}_{\phi}=\frac{Np_{\phi}^2}{4a^3}\frac{F'}{F^2},\\
\dot{T}=N,\\
\dot{p}_T=0\rightarrow p_T=\mathrm{const.}\,,
\end{array}
\right.
\end{eqnarray}
where $F'=\frac{dF(\phi)}{d\phi}$. Up to this point the cosmological
model, in view of the concerning issue of time, has been of course
under-determined. Before trying to solve these equations we must
decide on a choice of time in the theory. The under-determinacy
problem at the classical level may be resolved by using the gauge
freedom via fixing the gauge. A glance at the above equations shows
that choosing the gauge $N=1$, we have $\dot{T}=1\Rightarrow T=t$,
which means that variable $T$ may play the role of time in the
model. With this time gauge we obtain the following equation of
motion for $\phi$,
\begin{eqnarray}\label{SF}
2\frac{\ddot{\phi}}{\dot{\phi}}+\frac{F'}{F}\dot{\phi}+6\frac{\dot{a}}{a}=0.
\end{eqnarray}
This equation can easily be integrated to yield
\begin{eqnarray}\label{conservation_law}
F(\phi)\dot{\phi}^2=Ca^{-6},
\end{eqnarray}
where $C$ is an integration constant. Also, eliminating the momenta
from the system (\ref{classical dynamics1}) results

\begin{eqnarray}\label{constraint1}
\dot{a}^2+g_c-g_{\Lambda}a^2-\frac{g_r}{a^2}-\frac{g_s+C}{a^4}-\frac{p_T}{a}=0,
\end{eqnarray}in which we have used Eq.~(\ref{conservation_law}). In general, this
equation does not seem to have exact solution, so we restrict
ourselves to the especial case in which $g_c=g_{\Lambda}=g_r=0$,
$g_s\neq0$, for which the solution to Eq.~(\ref{constraint1}) reads
\begin{eqnarray}\label{1_a(t)_gs}
a(t)=\left(\frac{9p_T}{4}t^2-\frac{g_s+C}{p_T}\right)^\frac{1}{3}.
\end{eqnarray}
What remains to be found is an expression for the scalar field
$\phi(t)$. In the following, we shall consider the case of a
coupling function in the form $F(\phi)=\lambda \phi^m$. With this
choice for the function $F(\phi)$, and with the help of
Eqs.~(\ref{conservation_law}) and (\ref{1_a(t)_gs}) we are able to
calculate the time evolution of the scalar field as
\begin{eqnarray}\label{1_phi(t)_gs}
\phi(t)=\left[\phi_0-\frac{m+2}{6}\sqrt{\frac{C}{(g_s+C)\lambda}}\,
\ln\frac{3p_T t-2\sqrt{g_s+C}}{3p_T
t+2\sqrt{g_s+C}}\right]^{\frac{2}{m+2}},
\end{eqnarray}where $\phi_0$ is an integration constant and we
assumed $m\neq -2$. Finally, to understand the relation between the
big-bang singularity $a\rightarrow 0$ and the blow up singularity
$\phi\rightarrow \pm \infty$, we are going to find a classical
trajectory in configuration space $(a,\phi)$, where the time
parameter $t$ is eliminated. From (\ref{conservation_law}) and
(\ref{constraint1}) one gets
\begin{eqnarray}
\phi^m\left(\frac{d\phi}{da}\right)^2=\frac{C a^{-6}}{\lambda}
\left(-g_c+g_{\Lambda}a^2+\frac{g_r}{a^2}+\frac{g_s+C}{a^4}+\frac{p_T}{a}\right)^{-1},
\end{eqnarray}where for the case $g_c=g_{\Lambda}=g_r=0,
g_s\neq0$, after integration  reads
\begin{eqnarray}\label{1_phi(a)_gs}
\phi(a)=\left[\phi_0-\frac{m+2}{6}\sqrt{\frac{C}{(g_s+C)\lambda}}\,
\ln\frac{\sqrt{p_Ta^3+g_s+C}-\sqrt{g_s+C}}{\sqrt{p_Ta^3+g_s+C}+\sqrt{g_s+C}}\right]
^{\frac{2}{m+2}}.
\end{eqnarray}We see that the evolution of the
universe based on (\ref{1_a(t)_gs}) has big-bang-like singularities
at $t=\pm t_{*}$ where $t_{*}=\frac{2}{3P_T}\sqrt{g_s+C}$. Indeed,
the condition $a(t)\geq 0$ separates two sets of solutions each of
which is valid for $t\leq -t_{*}$ and $t\geq +t_{*}$, respectively.
For the former, we have a contracting universe which decreases its
size according to a power law relation and ends its evolution in a
singularity at $t=-t_{*}$, while for the latter, the evolution of
the universe begins with a big-bang singularity at $t=+t_{*}$ and
then follows the power law expansion $a(t)\sim t^{2/3}$ at late time
of cosmic evolution. On the other hand, the scalar field has a
monotonically decreasing behavior coming from $\phi\rightarrow
+\infty$ at early times and reaches to zero as time grows, see
Fig.~\ref{fig:figure_31}. We shall see in the next subsection how
this classical picture may be modified if one takes into account the
quantum mechanical considerations.

\subsection{The quantum model}
\begin{figure}[t]
\begin{center}
\includegraphics[width=0.6\textwidth]{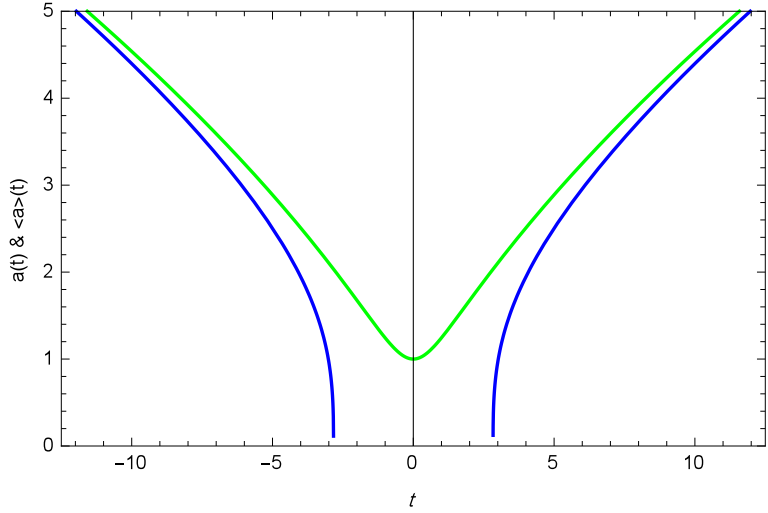}
\caption{The dynamical behavior of $\langle a\rangle(T)$ (green
line) in comparison with classical scale factor $a(t)$ (blue line).
See Eqs.~(\ref{1_expectationvalue_a_gs}) and (\ref{1_a(t)_gs}).}
\label{fig:figure_32}
\end{center}
\end{figure}

We now focus attention on the study of the quantum cosmology of the
model described above. We start by writing the Wheeler-DeWitt
equation from the Hamiltonian (\ref{hamiltonian1}). Since the lapse
function appears as a Lagrange multiplier in the Hamiltonian, we
have the Hamiltonian constraint $H=0$. Thus, application of the
Dirac quantization procedure demands that the quantum states of the
universe should be annihilated by the operator version of $H$, that
is $H\Psi(a,\phi,T)=0$, where $\Psi(a,\phi,T)$ is the wave function
of the universe. Use of the usual representation $P_q \rightarrow
-i\partial_q$ we are led to the following SWD equation
\begin{eqnarray}\label{SWD_eq_a_phi_T_1}
&&\frac{1}{4a}\left(\frac{\partial^2}{\partial a^2}
+\frac{\beta}{a}\frac{\partial}{\partial a}\right)\Psi(a,\phi,T)+
\left(-g_c a+g_{\Lambda}a^3+\frac{g_r}{a}+\frac{g_s}{a^3}\right)\Psi(a,\phi,T)\nonumber\\
&&-\frac{1}{4Fa^3}\left(\frac{\partial^2}{\partial\phi^2}
+\frac{\kappa
F'}{F}\frac{\partial}{\partial\phi}\right)\Psi(a,\phi,T)
=i\frac{\partial\Psi(a,\phi,T)}{\partial\ T},
\end{eqnarray}
where the parameters $\beta$ and $\kappa$ represent the ambiguity in
the ordering of factors $(a,P_a)$ and $(\phi,P_{\phi})$
respectively. This equation takes the form of a Schr\"{o}dinger
equation $i\partial \Psi/\partial T=H\Psi$, in which the Hamiltonian
operator is Hermitian with the standard inner product
\begin{eqnarray}\label{inner_product_1}
\langle\Psi_1\mid \Psi_2
\rangle=\int_{(a,\phi)}dad\phi\,a\;\Psi^*_1\Psi_2.
\end{eqnarray}
We separate the variables in above equation as
$\Psi(a,\phi,T)=e^{iET}\psi(a,\phi)$ leading to
\begin{eqnarray}\label{SWD_eq_a_phi_1}
&&\frac{1}{4a}\left(\frac{\partial^2}{\partial a^2}
+\frac{\beta}{a}\frac{\partial}{\partial a}\right)\psi(a,\phi)
-\frac{1}{4Fa^3}\left(\frac{\partial^2}{\partial\phi^2}
+\frac{\kappa F'}{F}\frac{\partial}{\partial\phi}\right)\psi(a,\phi)\nonumber\\
&&+\left(-g_c
a+g_{\Lambda}a^3+\frac{g_r}{a}+\frac{g_s}{a^3}+E\right)\psi(a,\phi)=0,
\end{eqnarray}
where $E$ is a separation constant. The solutions of the above
differential equation are separable and may be written in the form
$\psi(a,\phi)=A(a)\Phi(\phi)$ which yields
\begin{eqnarray}\label{SWD_eq_a_1}
&&\frac{d^2A(a)}{da^2}+\frac{\beta}{a}\frac{dA(a)}{da}
+4\left(-g_ca^2+g_{\Lambda}a^4+g_r+\frac{g_s+w}{a^2}+Ea\right)A(a)=0,\\\label{SWD_eq_phi}
&&\frac{d^2\Phi(\phi)}{d\phi^2} +\frac{\kappa
F'(\phi)}{F(\phi)}\frac{d\Phi(\phi)}{d\phi}+4wF(\phi)\Phi(\phi)=0,
\end{eqnarray}
where $w$ is another constant of separation. The factor-ordering
parameters does not affect the semiclassical probabilities
\cite{Hartle&Hawking}, so in what follows we have chosen $\beta=0$
and $\kappa=-1$ to make the differential equations solvable. Upon
substituting the relation $F(\phi)=\lambda \phi^m$ into
(\ref{SWD_eq_phi}), its solutions read in terms of the Bessel
functions $J$ and $Y$ as
\begin{eqnarray}\label{phi wave}
\Phi(\phi)=\,C_1\,\phi^\frac{1+m}{2}\,\mathrm{J}_{\frac{m+1}{m+2}}\left(\frac{4\sqrt{\lambda
w}}{m+2} \phi^{\frac{m+2}{2}}\right)+
\,C_2\,\phi^\frac{1+m}{2}\,\mathrm{Y}_{\frac{m+1}{m+2}}\left(\frac{4\sqrt{\lambda
w}}{m+2} \phi^{\frac{m+2}{2}}\right),
\end{eqnarray}
for $m\neq-2$ and
\begin{eqnarray}
\Phi(\phi)=\,C_1\,\phi^{\frac{-1+\sqrt{1-16\lambda w}}{2}}
+\,C_2\,\phi^{\frac{-1-\sqrt{1-16\lambda w}}{2}},
\end{eqnarray}
for $m=-2$. Also, if we set (as in the classical solutions)
$g_c=g_{\Lambda}=g_r=0$, Eq.~(\ref{SWD_eq_a_1}) admits the solution
\begin{eqnarray}\label{1_wavefunction_gs}
A(a)=c_1\sqrt{a}\,\mathrm{J}_{\nu}\left(\frac{4}{3}\sqrt{E}a^\frac{3}{2}\right)
+c_2\sqrt{a}\,\mathrm{Y}_{\nu}\left(\frac{4}{3}\sqrt{E}a^\frac{3}{2}\right),
\end{eqnarray}where $\nu=\frac{1}{3}\sqrt{1-16(g_s+w)}$. Thus, the eigenfunctions of the SWD equation can be written as
\begin{eqnarray}
\Psi_{E,w}(a,\phi,T)&=&e^{iET}A(a)\Phi(\phi)\nonumber\\\label{1_eigenfunctions_gs}
&=&e^{iET}\,\sqrt{a}\,\mathrm{J}_{\nu}\left(\frac{4}{3}\sqrt{E}a^{\frac{3}{2}}\right)\,
\phi^\frac{m+1}{2}\,\mathrm{J}_{\frac{m+1}{m+2}}\left(\frac{4\sqrt{\lambda
w}}{m+2}\phi^{\frac{m+2}{2}}\right),
\end{eqnarray}where we have chosen $C_2=c_2=0$ for having well-defined functions in all ranges of variables $a$ and
$\phi$. We may now write the general solutions to the SWD equations
as a superposition of the eigenfunctions, that is
\begin{eqnarray}
\Psi(a,\phi,T)&=&\int dE\,dw\,f(E)\,g(w)\,\Psi_{E,w}(a,\phi,T)\nonumber\\
&=&\sqrt{a}\phi^\frac{m+1}{2}\int_0^{w_0}dw\,g(w)\,\mathrm{J}_{\frac{m+1}{m+2}}
\left(\frac{4\sqrt{\lambda
w}}{m+2}\phi^{\frac{m+2}{2}}\right)\nonumber\\\label{1_wavepacket1_gs}
&&\qquad\times\int_0^{\infty}dE\,f(E)\,e^{iET}\,
\mathrm{J}_{\nu}\left(\frac{4}{3}\sqrt{E}a^{\frac{3}{2}}\right),
\end{eqnarray}
where $w_0=\frac{1}{16}-g_s$ and $f(E)$ and $g(w)$ are suitable
weight functions to construct the wave packets. By using the
equality \cite{book1}
\begin{eqnarray}\label{inttegral_a}
\int_0^{\infty}dx\,e^{-Zx^2}\,x^{\nu+1}\,\mathrm{J}_{\nu}(bx)=\frac{b^{\nu}}{(2Z)^{\nu+1}}
e^{-\frac{b^2}{4Z}},
\end{eqnarray}we can evaluate the integral over $E$ in (\ref{1_wavepacket1_gs}) and simple
analytical expression for this integral is found if we choose the
function $A(E)$ to be
\begin{eqnarray}
f(E)=E^{\frac{\nu}{2}}e^{-\sigma E},
\end{eqnarray}where $\sigma$ is an arbitrary positive constant. With
this procedure we get
\begin{eqnarray}
\Psi(a,\phi,T)&=&\sqrt{a}\,\phi^\frac{m+1}{2}\int_0^{w_0}dw\,g(w)\,
\mathrm{J}_{\frac{m+1}{m+2}}\left(\frac{4\sqrt{\lambda
w}}{m+2}\phi^{\frac{m+2}{2}}\right)\nonumber\\\label{1_wavepacket2_gs}
&&\qquad\times\frac{\left(\frac{4}{3}a^{\frac{3}{2}}\right)^{\frac{1}{3}\sqrt{1-16(g_s+w)}}}
{(2Z)^{1+\frac{1}{3}\sqrt{1-16(g_s+w)}}}e^{\frac{-4a^3}{9Z}},
\end{eqnarray}
where $Z=\sigma-iT$. To achieve an analytical closed expression for
the wave function, we assume that the above superposition is taken
over such values of $w$ for which one can use the approximation
$\sqrt{1-16(g_s+w)}\simeq\sqrt{1-16g_s}$, that is
\begin{eqnarray}
\Psi(a,\phi,T)&=&\sqrt{a}\,\phi^\frac{m+1}{2}\frac{\left(\frac{4}{3}a^{\frac{3}{2}}\right)
^{\frac{1}{3}\sqrt{1-16g_s}}}
{(2Z)^{1+\frac{1}{3}\sqrt{1-16g_s}}}e^{\frac{-4a^3}{9Z}}\nonumber\\\label{1_wavepacket3_gs}
&&\qquad\times\int_0^{w_0}dw\,g(w)\,\mathrm{J}_{\frac{m+1}{m+2}}\left(\frac{4\sqrt{\lambda
w}}{m+2} \phi^{\frac{m+2}{2}}\right).
\end{eqnarray}Now, by using the equality \cite{book1}
\begin{eqnarray}\label{inttegral_phi}
\int_0^1d\nu\,\nu^{r+1}(1-\nu^2)^{s/2}\mathrm{J}_{r}(z\nu)=\frac{2^s\,\Gamma(s+1)}{z^{s+1}}\mathrm{J}_{r+s+1}(z),
\end{eqnarray}and choosing  the weight
function $g(w)=\left(\frac{w}{w_0}\right)^{\frac{m+1}{2(m+2)}}\left(1-\frac{w}{w_0}\right)^{s/2}$, we are led to the following
expression for the wave function
\begin{eqnarray}
\Psi(a,\phi,T)&=&{\cal
N}\frac{a^{\frac{1+\sqrt{1-16g_s}}{2}}}{(\sigma-iT)^{1+\frac{1}{3}\sqrt{1-16g_s}}}
\exp\left(-\frac{4a^3}{9(\sigma-iT)}\right)\nonumber\\\label{1_wavepacket_gs}
&&\qquad\times\phi^{-\frac{1+(m+2)s}{2}}\mathrm{J}_{\frac{2m+3}{m+2}+s}
\left(\frac{\sqrt{(1-16g_s)\lambda}}{m+2}\phi^{\frac{m+2}{2}}\right),
\end{eqnarray}
where ${\cal N}$ is a normalization coefficient. Now, having the
above expression for the wave function  of the universe, we are
going to obtain the predictions for the behavior of the dynamical
variables in the corresponding cosmological model. In general, one
of the most important features in quantum cosmology is the recovery
of classical cosmology from the corresponding quantum model or, in
other words, how can the WD wave functions predict a classical
universe. In this approach, one usually constructs a coherent wave
packet with good asymptotic behavior in the minisuperspace, peaking
in the vicinity of the classical trajectory. On the other hand, in
an another approach to show the correlations between classical and
quantum pattern, following the many-worlds interpretation of quantum
mechanics, one may calculate the time dependence of the expectation
value of a dynamical variable $q$ as
\begin{equation}\label{AS}
\langle
q\rangle(t)=\frac{\langle\Psi|q|\Psi\rangle}{\langle\Psi|\Psi\rangle}.
\end{equation}
Following this approach, we may write the expectation value for the
scale factor as
\begin{eqnarray}
\langle
a\rangle(T)&=&\frac{\int_{a=0}^{\infty}\int_{\phi=-\infty}^{+\infty}
da\,d\phi\,a^2\,|\Psi|^2}
{\int_{a=0}^{\infty}\int_{\phi=-\infty}^{+\infty}
da\,d\phi\,a\,|\Psi|^2}\nonumber\\\label{1_expectationvalue_a_gs}
&=&\frac{3}{2}\frac{\Gamma\left(\frac{4+\sqrt{1-16g_s}}{3}\right)}
{\Gamma\left(\frac{3+\sqrt{1-16g_s}}{3}\right)}\left(\frac{\sigma^2+T^2}{3\sigma}\right)^{\frac{1}{3}}.
\end{eqnarray}It is important to
classify the nature of the quantum model as concerns the presence or
absence of singularities. For the wave function
(\ref{1_wavepacket_gs}), the expectation value
(\ref{1_expectationvalue_a_gs}) of $a$ never vanishes, showing that
these states are nonsingular. Indeed, the expression
(\ref{1_expectationvalue_a_gs}) represents a bouncing universe with
no singularity where its late time behavior coincides to the late
time behavior of the classical solution (\ref{1_a(t)_gs}), that is
$a(t)\sim t^{\frac{2}{3}}$. We have plotted this behavior in
Fig.~\ref{fig:figure_32}. As this figure shows instead of two
separate contracting and expanding classical solutions, the quantum
expectation value consists of two branches. In one branch the
universe contracts and when reaches a minimum size undergoes to an
expansion period. Therefore, we have bouncing cosmology in which the
bounce occurs at classical singularity. In a similar manner, the
expectation value for the scalar field reads as
\begin{eqnarray}\label{1_expectationvalue_phi_a_goto_0_gr=0}
\langle\phi\rangle(T)=\frac{\int da\,d\phi\,a\phi\,|\Psi|^2} {\int
da\,d\phi\,a\,|\Psi|^2}=\mathrm{const}.
\end{eqnarray}
We see that the expectation value of $\phi$ does not depend on time.
This result is comparable with those obtained in \cite{14-1} where a
constant expectation value for the dilatonic field in a quantum
cosmological model based on the string effective action coupled to
matter has been obtained.

\section{The $Sp_{\epsilon}^{\alpha+1}\ll Aa^{3(\alpha+1)}$ limit}\label{Sec_Limit2}
Now, let us return to the Hamiltonian (\ref{full_hamiltonian}) but
this time expand it in the late time limit
$Sp_{\epsilon}^{\alpha+1}\ll Aa^{3(\alpha+1)}$ as
\begin{eqnarray}
\left(Sp_{\epsilon}^{\alpha+1}+Aa^{3(\alpha+1)}\right)^{\frac{1}{\alpha+1}}&=&
A^{\frac{1}{\alpha+1}}a^3\left(1+\frac{Sp_{\epsilon}^{\alpha+1}}{Aa^{3(\alpha+1)}}\right)
^{\frac{1}{\alpha+1}}\nonumber\\
&=&A^{\frac{1}{\alpha+1}}a^3\Big[1+\frac{1}{\alpha+1}
\frac{Sp_{\epsilon}^{\alpha+1}}{Aa^{3(\alpha+1)}}\nonumber\\
&&\quad+\frac{1}{2}\frac{1}{\alpha+1}\left(\frac{1}{\alpha+1}-1\right)
\left(\frac{Sp_{\epsilon}^{\alpha+1}}{Aa^{3(\alpha+1)}}\right)^2+\ldots\Big]\nonumber\\\label{approximation2}
&\simeq&A^{\frac{1}{\alpha+1}}a^3+\frac{1}{\alpha+1}
\frac{A^{\frac{-\alpha}{\alpha+1}}Sp_{\epsilon}^{\alpha+1}}{a^{3\alpha}}.
\end{eqnarray}
Therefore, the super-Hamiltonian takes the form
\begin{eqnarray}\label{approximated_hamiltonian2}
H=N\,\left(-\frac{p_a^2}{4a}-g_ca+\bar{g}_{\Lambda}a^3+\frac{g_r}{a}+\frac{g_s}{a^3}
+\frac{p_{\phi}^2}{4F(\phi)a^3}+\frac{1}{\alpha+1}
\frac{A^{\frac{-\alpha}{\alpha+1}}Sp_{\epsilon}^{\alpha+1}}{a^{3\alpha}}\right),
\end{eqnarray}
where $\bar{g}_{\Lambda}=g_{\Lambda}+A^{\frac{1}{\alpha+1}}$. Now,
consider the following canonical transformation
\cite{Pedram&Jalalzadeh}
\begin{eqnarray}\label{canonical_transformation2}
\begin{array}{ll}
&T=-(\alpha+1)A^{\frac{\alpha}{\alpha+1}}p_{\epsilon}^{-(\alpha+1)}p_S,\\
&p_T=\frac{1}{\alpha+1}A^{\frac{-\alpha}{\alpha+1}}Sp_{\epsilon}^{\alpha+1},
\end{array}
\end{eqnarray}
under act of which the above Hamiltonian becomes
\begin{eqnarray}\label{hamiltonian2}
H=N\,\left(-\frac{p_a^2}{4a}-g_ca+\bar{g}_{\Lambda}a^3+\frac{g_r}{a}+\frac{g_s}{a^3}
+\frac{p_{\phi}^2}{4F(\phi)a^3}+\frac{p_T}{a^{3\alpha}}\right).
\end{eqnarray}We now may repeat the steps as we have taken in the
previous section to obtain the classical and quantum cosmological
dynamics based on the Hamiltonian (\ref{hamiltonian2}).

\subsection{The classical model}
By the Hamiltonian (\ref{hamiltonian2}) the classical equations of
motion are
\begin{eqnarray}\label{classical dynamics2}
\left\{
\begin{array}{ll}
\dot{a}=-\frac{Np_a}{2a},\\
\dot{p}_a=N\left(-\frac{p_a^2}{4a^2}+g_c-3\bar{g}_{\Lambda}a^2+\frac{g_r}{a^2}+\frac{3g_s}{a^4}
+\frac{3p_{\phi}^2}{4Fa^4}+\frac{3\alpha p_T}{a^{3\alpha+1}}\right),\\
\dot{\phi}=\frac{Np_{\phi}}{2Fa^3},\\
\dot{p}_{\phi}=\frac{Np_{\phi}^2}{4a^3}\frac{F'}{F^2},\\
\dot{T}=\frac{N}{a^{3\alpha}},\\
\dot{p}_T=0\rightarrow p_T=\mathrm{const.}
\end{array}
\right.
\end{eqnarray}
To have the clock parameter as $T=t$, we should choose the lapse
function $N=a^{3\alpha}$. Since the third and the fourth equations
of this system are the same as their counterparts in the system
(\ref{classical dynamics1}), the dynamical equations for the scalar
field are the same as Eqs.~(\ref{SF}) and (\ref{conservation_law}).
Also, with the constraint equation $H=0$ we obtain
\begin{eqnarray}\label{constraint2}
\dot{a}^2+a^{6\alpha}\left(g_c-\bar{g}_{\Lambda}a^2-\frac{g_r}{a^2}-\frac{g_s+C}{a^4}
-\frac{p_T}{a^{3\alpha+1}}\right)=0.
\end{eqnarray}
To solve this equation we suppose $g_c=\bar{g}_{\Lambda}=0$ and
$g_r,g_s\neq 0$ which simplifies the above equation as
\begin{eqnarray}\label{constraint2_grgs}
\dot{a}^2=a^{6\alpha}\left(\frac{g_r}{a^2}+\frac{g_s+C}{a^4}+\frac{p_T}{a^{3\alpha+1}}\right).
\end{eqnarray}This equation does not yet have exact solution for
general case with arbitrary $\alpha$. So, from now on we restrict
ourselves to the case $\alpha=\frac{1}{3}$ for which the solution to
Eq.~(\ref{constraint2_grgs}) is
\begin{eqnarray}\label{2_a(t)_grgs}
a(t)=\sqrt{(g_r+p_T)t^2-\frac{g_s+C}{g_r+p_T}}.
\end{eqnarray}By means of this relation, with the help of (\ref{SF}) and
(\ref{conservation_law}) and with the same detail as in previous
section, we get the following expressions for $\phi(t)$ and
$\phi(a)$

\begin{eqnarray}\label{phi(t)_grgs}
\phi(t)=\left[\phi_0-\frac{m+2}{4}\sqrt{\frac{C}{(g_s+C)\lambda}}\,
\ln\frac{(g_r+p_T)t-\sqrt{g_s+C}}{(g_r+p_T)t+\sqrt{g_s+C}}\right]^{\frac{2}{m+2}},
\end{eqnarray}
and
\begin{eqnarray}\label{phi(a)_grgs}
\phi(a)=\left[\phi_0+\frac{m+2}{2}\sqrt{\frac{C}{(g_s+C)\lambda}}
\ln\frac{\sqrt{g_s+C}+\sqrt{(g_r+p_T)a^2+g_s+C}}{a}\right]^{\frac{2}{m+2}}.
\end{eqnarray}

\subsection{The quantum model}
\begin{figure}[t]
\begin{center}
\includegraphics[width=0.6\textwidth]{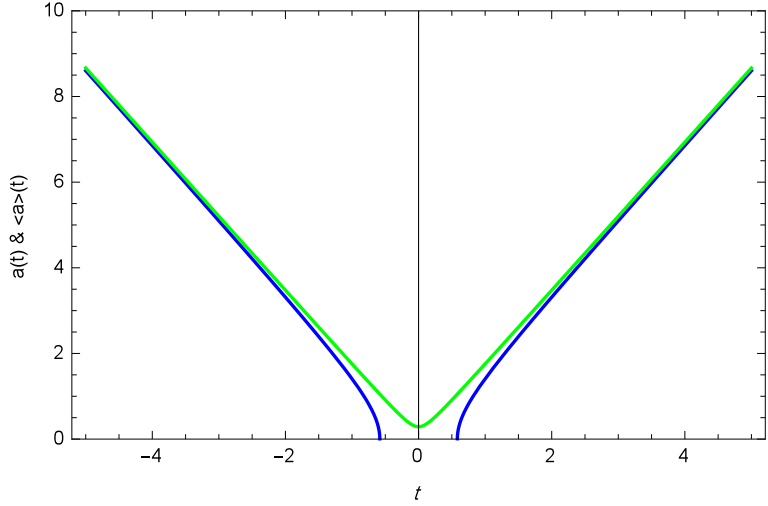}
\caption{Qualitative behavior of $a(t)$ (blue line) and $\langle
a\rangle(t)$ (green line), see Eqs.~(\ref{2_a(t)_grgs}) and
(\ref{2_expectationvalue_a_grgs}).} \label{fig:figure_42}
\end{center}
\end{figure}

The standard quantization process based on the Hamiltonian
(\ref{hamiltonian2}) get us the following SWD equation
\begin{eqnarray}\label{SWD_eq_a_phi_T_2}
&&\frac{1}{4a}\left(\frac{\partial^2}{\partial a^2}
+\frac{\beta}{a}\frac{\partial}{\partial a}\right)\Psi(a,\phi,T)+
\left(-g_c a+\bar{g}_{\Lambda}a^3+\frac{g_r}{a}+\frac{g_s}{a^3}\right)\Psi(a,\phi,T)\nonumber\\
&&-\frac{1}{4Fa^3}\left(\frac{\partial^2}{\partial\phi^2}
+\frac{\kappa
F'}{F}\frac{\partial}{\partial\phi}\right)\Psi(a,\phi,T)
=\frac{i}{a^{3\alpha}}\frac{\partial\Psi(a,\phi,T)}{\partial\ T},
\end{eqnarray}
where $\beta$ and $\kappa$ are again factor ordering parameters
which as before we set them as $\beta=0$ and $\kappa=-1$. This time
the Hamiltonian operator is Hermitian with the inner product
\begin{eqnarray}\label{inner_product_2}
\langle\Psi_1,\Psi_2\rangle=\int_{(a,\phi)}dad\phi\,a^{1-3\alpha}\;\Psi^*_1\Psi_2.
\end{eqnarray}Separation of variables as
$\Psi(a,\phi,T)=e^{iET}A(a)\Phi(\phi)$ lead to
Eq.~(\ref{SWD_eq_phi}) with solution (\ref{phi wave}) for the
$\phi$-sector of the eigenfunctions while for $A(a)$ we are arrived
at the following equation (with $g_c=\bar{g}_{\Lambda}=0$)
\begin{eqnarray}\label{2_SWD_eq_grgs}
\frac{d^2A}{da^2}+4\left(g_r+\frac{g_s+w}{a^2}+\frac{E}{a^{3\alpha-1}}\right)A=0.
\end{eqnarray}
For $\alpha=\frac{1}{3}$ this equation has the solutions
\begin{eqnarray}\label{2_wavefunction_a_grgs}
A(a)=c_1\sqrt{a}\,\mathrm{J}_{\nu}(2\sqrt{g_r+E}a)+c_2\sqrt{a}\,\mathrm{Y}_{\nu}(2\sqrt{g_r+E}a),
\end{eqnarray}with $\nu=\frac{1}{2}\sqrt{1-16(g_s+w)}$.
Therefore, the eigenfunctions of the corresponding SWD equation read
\begin{eqnarray}\label{2_eigenfunctions_grgs}
\Psi_{E,w}(a,\phi,T)=e^{iET}\,\sqrt{a}\,\mathrm{J}_{\nu}\left(2\sqrt{g_r+E}a\right)\,\phi^\frac{m+1}{2}\,
\mathrm{J}_{\frac{m+1}{m+2}}\left(\frac{4\sqrt{\lambda
w}}{m+2}\phi^{\frac{m+2}{2}}\right),
\end{eqnarray}in which we have removed the Bessel functions $Y$ from
the solutions. Following the same steps which led us to the wave
function (\ref{1_wavepacket_gs}), we obtain the wave function as
\begin{eqnarray}
\Psi(a,\phi,T)&=&{\cal
N}\,e^{-ig_rT}\,\frac{a^{\frac{1+\sqrt{1-16g_s}}{2}}}
{(\sigma-iT)^{1+\frac{1}{2}\sqrt{1-16g_s}}}
\exp\left(\frac{-a^2}{\sigma-iT}\right)\nonumber\\\label{1_wavepacket_grgs}
&&\qquad\times\phi^{-\frac{1+(m+2)s}{2}}\mathrm{J}_{\frac{2m+3}{m+2}+s}
\left(\frac{\sqrt{(1-16g_s)\lambda}}{m+2}\phi^{\frac{m+2}{2}}\right),
\end{eqnarray}from which the expectation values are obtained as
\begin{eqnarray}
\langle a\rangle(T)&=&\frac{\int da\,d\phi\,a\,|\Psi|^2} {\int
da\,d\phi\,|\Psi|^2}\nonumber\\\label{2_expectationvalue_a_grgs}
&=&\frac{\Gamma\left(\frac{3+\sqrt{1-16g_s}}{2}\right)}
{\Gamma\left(\frac{2+\sqrt{1-16g_s}}{2}\right)}
\left(\frac{\sigma^2+T^2}{2\sigma}\right)^{\frac{1}{2}},\\\label{2_expectationvalue_phi_grgs}
\langle\phi\rangle(T)&=&\frac{\int da\,d\phi\,\phi\,|\Psi|^2} {\int
da\,d\phi\,|\Psi|^2}=\mathrm{const.}
\end{eqnarray}
In Fig.~\ref{fig:figure_42} we have plotted the
classical scale factor (\ref{2_a(t)_grgs}) and its quantum
expectation value (\ref{2_expectationvalue_a_grgs}). The discussions
on the comparison between quantum cosmological solutions and their
corresponding form from the classical formalism, are the same as
previous section. Similar discussion as above would be applicable to
this case as well.

\section{Conclusion}\label{sec_conclusion}
In this paper we have applied the Ho\v{r}ava theory of gravity to a
FRW cosmological model coupled minimally to a scalar field in which
a generalized Chaplygin gas, in the context of the Schutz'
representation, plays the roll of the matter field. The use of the
Schutz' formalism for Chaplygin gas allowed us to introduce the only
remaining matter degree of freedom as a time parameter in the model.
After a very brief review of HL theory of gravity, we have
considered a FRW cosmological setting in the framework of the
projectable HL gravity without detailed balance condition and
presented its Hamiltonian in terms of the minisuperspace variables.
Though the corresponding classical equations did not have exact
solutions, we analyzed their behavior in the limiting cases of the
early and late times of cosmic evolution and obtained analytical
expressions for the scale factor and the scalar field in these
regions. We have seen that these solutions are consisted of two
separate branches each of which exhibit some kinds of classical
singularities. Indeed, the classical solutions have either
contracting or expanding branches which are disconnected from each
other by some classically forbidden regions. Another part of the
paper is devoted to the quantization of the model described above in
which we saw that the classical singular behavior will be modified.
In the quantum models, we showed that the SWD equation can be
separated and its eigenfunctions can be obtained in terms of
analytical functions. By an appropriate superposition of the
eigenfunctions, we constructed the corresponding wave packets. Using
Schutz's representation for the Chaplygin gas, under a particular
gauge choice, we led to the identification of a time parameter which
allowed us to study the time evolution of the resulting wave
function. Investigation of the expectation value of the scale factor
shows a bouncing behavior near the classical singularity. In
addition to singularity avoidance, the appearance of bounce in the
quantum model is also interesting in its nature due to prediction of
a minimal size for the corresponding universe. It is well-known that
the idea of existence of a minimal length in nature is supported by
almost all candidates of quantum gravity. \vspace{5mm}\newline
\noindent {\bf Acknowledgement}\vspace{2mm}\noindent\newline The
research of P. Pedram is supported by the Iran National Science
Foundation (INSF), Grant No.~93047987.

\end{document}